\documentclass[12pt]{iopart}
\usepackage{graphicx}
\expandafter\let\csname equation*\endcsname\relax
\expandafter\let\csname endequation*\endcsname\relax
\usepackage{amssymb, amsmath}
\usepackage{bm,hyperref}
\begin{document}

\title[Control of conditional quantum beats]{Control of conditional quantum beats in cavity QED: amplitude decoherence and phase shifts.}

\author{A D Cimmarusti, C A Schroeder, B D Patterson and L A Orozco}

\address{Joint Quantum Institute, Department of Physics, University of
  Maryland and National Institute of Standards and Technology, College
  Park, MD 20742, USA}

 \author{P Barberis-Blostein}

\address{Instituto de Investigaciones en Matem\'aticas Aplicadas y en Sistemas, Universidad Nacional Aut\'onoma de M\'exico, Ciudad Universitaria, 04510, M\'exico, DF, M\'exico}

\author{H J Carmichael}

\address{Department of Physics, University of Auckland, Private Bag
  92019, Auckland,\\ New Zealand}

\ead{lorozco@umd.edu}
\begin{abstract}
We implement a simple feedback mechanism on a two-mode cavity QED system to preserve the Zeeman coherence of a ground state superposition that generates quantum beats on the second-order correlation function. Our investigation includes theoretical and experimental studies that show how to prevent a shift away from the Larmor frequency and associated decoherence caused by Rayleigh scattering. The protocol consists of turning off the drive of the system after the detection of a first photon and letting it evolve in the dark. Turning the drive back on after a pre-set time reveals a phase accumulated only from Larmor precession, with the amplitude of the quantum beat more than a factor of two larger than with continuous drive.

\end{abstract}

\pacs{42.50.Pq,42.50.Md,03.67.Ac}
\maketitle

\section{Introduction}
\label{sec:intro}
Preservation of quantum coherence is of fundamental importance in many fields, from atomic clocks to quantum information science. The tension between interaction with an environment to extract information and dissipation is at the heart of quantum open systems \cite{carmichaelbookv1,carmichaelbookv2}; attempts to isolate a system usually remove the possibility of measuring and controling its dynamics. Recent developments in quantum feedback \cite{wisemanbook,deutsch10,barberis10} and its application in quantum optics, however, point to an era where the theoretical tools and experimental time scales needed for control are within reach.

We work with an optical cavity QED system in the intermediate coupling regime. We profit from the internal structure of the atoms, which allows us look at ground state quantum coherences that outlive the atomic excited state and cavity lifetimes by orders of magnitude \cite{norris10}. These coherences are nevertheless rather delicate and suffer from phase and amplitude modifications as they are probed \cite{norris12}. In this paper we show experimentally and theoretically that it is possible to preserve the coherences, recovering both amplitude and phase, by following a protocol that starts with the detection of a photon, which then triggers a pulse to turn off the system drive.

This work goes beyond our previous experiments on quantum feedback in optical cavity QED \cite{smith02,reiner04a} where only the amplitude was recovered, without control over the phase. Moreover, that protocol depended critically on the specific time of feedback application after a photon detection. Our current work shares with it a reliance on strong quantum feedback, where we draw on our knowledge of the conditional dynamics of the system to capture (store) and at a later time release a quantum state.

More recent experiments aim at deterministic quantum control \cite{wilk07,weber09}. In contrast, we use spontaneous emission to prepare and detect the ground-state coherence. This renders our control protocol non-deterministic. The implementation of our fast feedback helps us go further in the context of studying the effects of drive duration and strength on the coherence and accumulated phase of our superposition.

The protocol is rather simple. It requires no processing but simply follows from the setting of a quantum beat phase by the detection of a single photon. This is in contrast with two recent quantum control studies. The first, recent experiments with Rydberg atoms in superconducting cavities \cite{sayrin11,zhou12}, performs extensive calculations based on measurement outcomes to create and maintain a Fock state in a microwave cavity. The second reports experiments and theory aiming for quantum control of the full ground state manifold of Cs \cite{smith06,merkel08,mischuck12}.

The paper is organized as follows. Section~\ref{sec:quant_beat} introduces the conditional ground-state quantum beats. Section~\ref{sec:apparatus} explains our experimental setup and data processing techniques. The theoretical treatment follows in Sec.~\ref{sec:theory} and the experimental results in Sec.~\ref{sec:results}. Section~\ref{sec:discussion} compares data with Monte-Carlo quantum trajectory simulations. The paper finishes with concluding remarks in Sec.~\ref{sec:conclusion}.

\section{Conditional quantum beats in a cavity QED system}
\label{sec:quant_beat}
An optical  cavity QED system isolates the interaction between atoms and a few modes of the electromagnetic field of a cavity. It allows the study of non-classical effects in the transmitted light that escapes from the cavity. Although the basic interaction is between the induced dipole of the atomic medium and the modes of the cavity, the setup also allows for the preparation of different states. Our recent work focuses on the long-lived coherences created by spontaneous emission in the ground state~\cite{norris10}. We describe in this section our system and its ground state superpositions.

\begin{figure}
\centering
\includegraphics[width=0.65\linewidth]{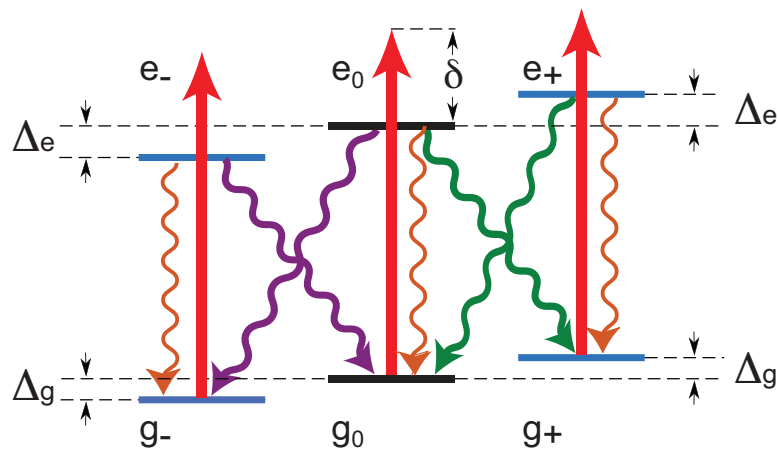}
\caption{Simplified atomic energy level structure of the $F=3 \rightarrow F=4$ $D_2$ line in $^{85}$Rb with Zeeman shifts. Different magnetic $g$-factors yield $\Delta_e > \Delta_g$. Both the $\pi$ (red and orange) and $\sigma$ (purple and green) transitions are indicated. We consider only situations with a $\pi$ drive, which might be detuned by $\delta$.}\label{fig:simplemodel}
\end{figure}

Our cold atomic beam interacts with the two orthogonal modes of a high finesse optical cavity. The $^{85}$Rb atoms exhibit ground- and excited-state Zeeman structure on the $D_2$ line with different magnetic $g$-factors (see Fig.~\ref{fig:simplemodel}). The laser drives $\pi$ (V polarization) transitions, $F=3, m \rightarrow F=4, m$, as indicated by the red arrows in the figure. Atomic excitation and decay transfer some of this energy to the orthogonal mode (H polarization). Spontaneous emission generates a long-lived Zeeman superposition in the ground-state (purple and green wavy lines). Its signature is a quantum beat seen in a conditional intensity measurement of the undriven mode. Using the simplified schematic of Fig.~\ref{fig:simplemodel}, if an atom enters the cavity in the $m=0$ ground state, the detection of a photon in the orthogonal mode sets the atom in a superposition of $m=\pm$ ground states. The prepared superposition then evolves in the magnetic field, acquiring a relative phase, until another $\pi$ excitation transfers the developed ground-state coherence to the excited state; subsequently, detecting a second (H polarized) photon projects the atom back into its starting state. The sequence overall realizes a quantum eraser \cite{scully82,zajonc83} as the intermediate ground-state is not observed.

Before emitting the second H polarized photon---a $\sigma$ transition---several intervening $\pi$ spontaneous emissions (Rayleigh scattering) can occur (orange wavy lines) \cite{uys10}. Each of these quantum jumps interrupts the atomic dipole and causes a small phase advance on the ground-state coherence, which accumulates over time to become a frequency shift~\cite{norris12}. The accumulated jumps occur randomly in time, so the frequency shift is accompanied by phase diffusion. Eventually, the diffusion dephases the coherence. Of course, a $\sigma$ spontaneous emission destroys the coherence all in one go for the simplified level structure in Fig.~\ref{fig:simplemodel}; not, however, for the considered transition in $^{85}$Rb, where there are actually seven rather then three Zeeman sublevels in the ground state.

The diagnostic tool used to understand and modify this spontaneous creation and evolution of ground-state coherence is a conditional measurement of the H-polarized (undriven) mode intensity, i.e., the second-order correlation function $g^{(2)}(\tau)$ of the H-polarized light. Two indistinguishable paths yield ``start'' and ``stop'' photons for the measurement: $|g_0\rangle\rightarrow|e_0\rangle\rightarrow |g_+\rangle\rightarrow|e_+\rangle \rightarrow |g_0\rangle$ and $|g_0\rangle\rightarrow|e_0\rangle\rightarrow|g_{-}\rangle\rightarrow|e_{-}\rangle\rightarrow|g_0\rangle$. Since the phase advance along each path is different in sign, though equal in magnitude, and the magnitude grows in time, interference between the paths yields oscillations: $g^{(2)}(\tau)$ exhibits a quantum beat.

\section{Apparatus}
\label{sec:apparatus}
This section explains our apparatus, data taking, data processing,  and the experimental realization of the feedback mechanism to protect the coherence of the ground-state superposition.

\begin{figure}
\centering
\includegraphics[width=0.95\linewidth]{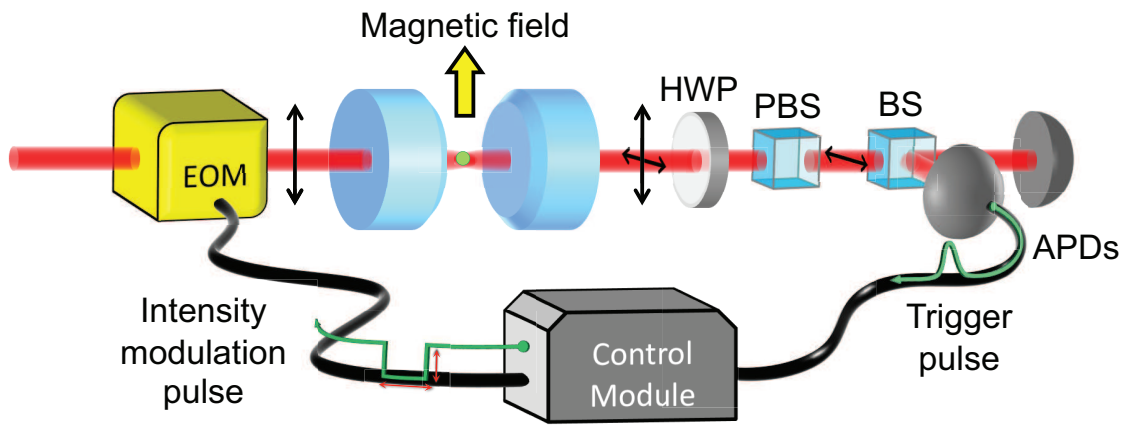}
\caption{Schematic of the experimental apparatus. The detection of a photon generates an electronic pulse that changes the amplitude of the laser drive for a pre-set amount of time. An electro optical modulator (EOM) sets the drive intensity. The light exits the cavity and passes through a half wave plate (HWP), a polarizing beam splitter (PBS), and a beam splitter (BS), which direct photons onto a pair of avalanche photodiodes (APDs). The photo-pulses from the APDs are correlated aginst the initial photon detection to obtain $g^{(2)}(\tau)$ (corrrelator not shown).}
 \label{fig:diagram}
\end{figure}

Figure \ref{fig:diagram} presents a schematic of the apparatus. Our optical cavity QED system is in the regime of intermediate coupling, where the dipole coupling constant ($g/2\pi=1.2 \times 10^6\mkern2mu{\rm s}^{-1}$) is comparable to the cavity and spontaneous emission decay rates ($\kappa/2 \pi=3 \times 10^{6}\mkern2mu{\rm s}^{-1}$ and $\gamma/2\pi=6 \times 10^6\mkern2mu{\rm s}^{-1}$). Our experiment consists of a $2\mkern2mu{\rm mm}$ Fabry-Perot cavity and a source of cold $^{85}$Rb atoms \cite{norris12b}. The source delivers, on average, a few maximally coupled atoms within the mode volume of the cavity at all times. This continuous cold atomic beam comes from an unbalanced Magneto-Optical Trap, a technique known as LVIS (Low Velocity Intense Source) \cite{lu96}. The cavity supports two degenerate modes of orthogonal linear polarization (H and V). During their $5\mkern2mu\mu{\rm s}$ transit, the atoms interact with the orthogonally polarized modes and can spontaneously emit into the cavity. We drive the $D_2$ line of $^{85}$Rb between the ground level $F=3$ and the excited level $F=4$ in the presence of a magnetic field of about 5 Gauss.

Birefringence from the cavity mirrors, vacuum chamber windows and lenses has a small effect on the frequency separation of the H and V modes. At the working intensities of the experiment the two modes are degenerate to better than $0.1\kappa$ and the extinction ratio at the output is better than $5 \times 10^{-4}$, a negligible contribution.

The light at $780\mkern2mu{\rm nm}$ passes through an EOSpace fiber electro-optic modulator before reaching the cavity. This device generates amplitude modulation sidebands at $227\mkern2mu{\rm  MHz}$ on the light. The upper sideband acts as the drive of the system and the carrier and the lower sideband reflect back from the cavity. The setup allows us to rapidly manipulate the amplitude of the drive. We use an SRS digital delay generator DG645 to generate an electronic pulse (risetime less than 10 ns) that opens a minicircuits ZAD-1-1 double balanced mixer, operating as an RF attenuator, to generate the $227\mkern2mu{\rm MHz}$ RF frequency that feeds the EOM. The output of the APDs (SPCM-AQR Perkin Elmer), in the form of TTL pulses, travels to a correlator card (Becker \& Hickl DPC-230), where each pulse arrival is time stamped and stored. We split one of the APD TTL pulses before reaching the correlator card and use it to trigger the DG645, which then drives the mixer. We set the length of the pulse and its amplitude with the DG645. The intrinsic electronic delay of the system is $325\mkern2mu{\rm ns}$, limited by the internal delay of the DG645 between external trigger and gate output.

The process of random photon emission via spontaneous decay translates into a stream of TTL pulses, which causes the DG645 to miss some triggers. The device can handle trigger rates up to $10\mkern2mu{\rm MHz}$, but from a synchronous source. Our photon detection rate of $\sim20\mkern2mu{\rm kHz}$ (start APD) causes about 2\% of missing triggers. We only keep the photon arrivals that successfully trigger the DG645 by implementing a software filter when processing the data. The DG645 produces a copy that we also send to the correlator card and use for the software filter~\footnote{auto-correlation and filtering computer code available at: \url{http://hdl.handle.net/1903/13306}}.

We detect the coherence using a homodyne process where a small part of the light exiting the drive mode is mixed with the signal using a half wave plate (HWP in Fig.~\ref{fig:diagram}). The presence of this allows us to look directly at the Larmor frequency in the Zeeman ground state superposition, and the oscillations show the interference of two fields, one that has the oscillation and one that does not.

The loss of coherence of the superposition is a degradation that is intimately related to the interrogation by the drive laser \cite{norris12}. The protocol that we present in this paper to eliminate it consists of reducing the amplitude of the $\pi$ drive to the cavity after a pre-set time from the detection of the first photon, and then returning the amplitude to the previous level after a fixed time to look at the oscillations. Since the frequency of oscillation is, to first order, set by the Larmor precession frequency; the atoms preserve the phase without interrogation, continuing their oscillation in the ground state.

\section{Theory}
\label{sec:theory}
Decoherence of the ground-state superposition arises from several factors. An important contribution comes from the dephasing process due to random Rayleigh scattering. In this section, we present --based on the phase shifts from quantum jumps-- a simplified model of the rate of decoherence due to this process and the associated shift in the Larmor precession frequency~\cite{norris12}.

A $\pi$ polarized coherent field with amplitude $\alpha$, resonant with the transition $|g_0\rangle\rightarrow|e_0\rangle$, drives $\pi$ transitions on the vertical mode of a cavity QED system in the presence of a magnetic field (see Fig.~\ref{fig:simplemodel}). The cavity provides two orthogonal modes for drive and detection of the system. Some of the spontaneous emission enters the orthogonal polarization mode $H$. The input to the correlator consists of the $H$ spontaneous emission mixed with a local oscillator of strength $\epsilon$. The detection of a first $H$ photon prepares a ground-state superposition \cite{norris10} that evolves in time as
\begin{equation}
  |\psi_g(t)\rangle = C_0\frac{1}{\sqrt{2}}(e^{i (\Delta_g+\Delta_{\rm AC})t}|g_-\rangle + e^{-i (\Delta_g+\Delta_{\rm AC})t}|g_+\rangle )
  +C_1 |g_0\rangle
\label{eq:groundstate}
\end{equation}
where the amplitudes $C_0$ and $C_1$ depend on the strength and phase of the local oscillator. The term $|g_0\rangle$ appears because it is not
possible to know the origin of the first detected photon in the presence of the local oscillator: it can come either from the local oscillator or from the atomic spontaneous emission. To lowest order in $g^2|\alpha|^2$ (drive intensity) the ground-state AC Stark shifts are
\begin{equation}
  \Delta_{\rm AC}= - \frac{g^2|\alpha|^2\Delta}{(\gamma/2)^2+\Delta^2}
\label{eq:ACStarkshift}
\end{equation}
for state $|g_+ \rangle$ and $-\Delta_{\rm AC}$ for $|g_- \rangle$.

The amplitudes in Eq.~(\ref{eq:groundstate}) couple to the corresponding excited-state amplitudes, driving a steady-state superposition:
\begin{equation}
  \label{eq:excited-jump}
  |\psi_e(t)\rangle = C_0\frac{g \alpha}{\sqrt{2}}\left(\frac{e^{i(\Delta_g+\Delta_{\rm AC})t}}{\gamma/2 - i \Delta} |e_-\rangle
    + \frac{e^{-i (\Delta_g+\Delta_{\rm AC})t}}{\gamma/2 + i \Delta} |e_+\rangle\right)+C_1\frac{g\alpha}{\gamma/2}|e_0\rangle.
\end{equation}
The excited-state amplitudes follow the ground-state oscillation; the excited-state splitting enters through the factors $\gamma/2\pm i \Delta$ only, which carry a phase shift. Consider now the effect of quantum jumps from spontaneous emission occurring during the interval in between the detection of a pair of $H$ polarized photons from the cavity, i.e., the $\pi$ jumps which constitute Rayleigh scattering in Fig.~\ref{fig:simplemodel}. With
jump rate $\Gamma=2g^2|\alpha|^2/(\gamma/2)$, the driven dipole between ground and excited states turns off and the amplitudes of Eq.~(\ref{eq:excited-jump}) are transferred to the ground state. It follows that each time a quantum jump occurs there is a phase advance; if $n$ quantum jumps occur, Eq.~(\ref{eq:groundstate}) is replaced by:

\begin{eqnarray}
\mkern-60mu{\cal{N}}_n|\psi_g(t)\rangle&=& \frac{C_0(\gamma/2)^n}{\sqrt2}\left\{\frac{(\gamma/2+i\Delta)^n}{[(\gamma/2)^2 + \Delta^2]^{n/2}} e^{i(\Delta_g+\Delta_{\rm AC})t}|g_-\rangle\right.\nonumber\\
&&\left.+\frac{(\gamma/2-i\Delta)^n}{[(\gamma/2)^2+\Delta^2]^{n/2}}e^{-i(\Delta_g+\Delta_{\rm AC})t}
|g_+\rangle\right\}+C_1[(\gamma/2)^2 + \Delta^2]^{n/2}|g_0\rangle\, ,
\label{ground-jump}
\end{eqnarray}
with normalization factor:
\begin{equation}
{\cal{N}}_n=\sqrt{|C_0|^2(\gamma/2)^{2n}+|C_1|^2[(\gamma/2)^2+\Delta^2]^n}.
\end{equation}
The ground-state superposition has acquired a phase advance. The number of quantum jumps increases with time, so the phase advance accumulates over time.

We average against a Poisson distribution with mean $\Gamma t$ to obtain the expectation value of the ground-state coherences for jump rate $\Gamma$:
\begin{eqnarray}
  \label{eq:coherence}
\rho_{g_{+},g_{-}} & =& e^{-2i(\Delta_g+\Delta_{\rm AC})t}\frac{|C_0|^2}{2}\sum_{n=0}^\infty\frac{(\gamma/2)^{2n}}{{\cal{N}}_n^2}\frac{(\gamma/2-i\Delta)^{2n}}{[(\gamma/2)^2 + \Delta^2]^{n}}
\frac{(\Gamma t)^n}{n!}e^{-\Gamma t}\nonumber\\
&\approx& e^{-2i(\Delta_g+\Delta_{\rm AC})t}\frac{|C_0|^2}{2}e^{-(2\Gamma_{\rm decoh}+i2\Delta_{\rm jump})t}
\end{eqnarray}
and
\begin{equation}
\label{eq:coherence2}
\rho_{g_{\pm},g_{0}}\approx e^{\mp i(\Delta_g+\Delta_{\rm AC})t}C_0^*C_1e^{-(\Gamma_{\rm decoh}\pm i\Delta_{\rm jump})t},
\end{equation}
where we assume $2\Delta/\gamma\ll1$ and take $(\gamma/2)^{2n}/{\cal{N}}_n^2\approx1$.

The imaginary part of the exponents in Eqs.~(\ref{eq:coherence}) and (\ref{eq:coherence2}) contains terms $-2\Delta_{\rm jump}t$ and $\mp\Delta_{\rm jump}t$, where to first order in $2\Delta/\gamma$:
\begin{equation}
\Delta_{\rm jump} = \Gamma \frac{2\Delta}{\gamma}=\frac{8 g^2|\alpha|^2\Delta}{\gamma^2}=-2\Delta_{\rm AC}.
\label{eq:jumpshift}
\end{equation}
These terms represent an additional frequency shift arising from the mean rate of phase accumulation from quantum jumps due to Rayleigh scattering. For the $(g_\pm,g_0)$-coherence, the net differential ground-state light shift, in the low drive limit, becomes:
\begin{equation}
\Delta_{\rm light}=(\Delta_{\rm AC}+\Delta_{\rm jump})\approx-\Delta_{\rm AC},
\label{eq:deltalight}
\end{equation}
and $2\Delta_{\rm light}=-2\Delta_{\rm AC}$ for the $(g_+,g_-)$-coherence.

The exponent in Eqs.~(\ref{eq:coherence}) and (\ref{eq:coherence2}) also contains a damping term, which decoheres the quantum beats at a rate (to lowest order in $2\Delta/\gamma$)
\begin{equation}
\Gamma_{\rm decoh} = \Gamma \frac{\Delta^2}{(\gamma/2)^2}=2g^2|\alpha|^2\frac{\Delta^2}{(\gamma/2)^3}.
  \label{eq:jumpwidth}
\end{equation}
The decoherence arises from the phase diffusion which accompanies the average phase drift responsible for the frequency shift. The two aspects, drift and diffusion, come together as a package from the stochastic nature of the jump process.

The probability of detecting a second photon with $H$ polarization following a trigger detection is proportional to
\begin{equation}
\langle \psi(\tau)|(\epsilon^*+\sigma^\dagger_{\rm H})(\epsilon+\sigma_{\rm H})|\psi(\tau)\rangle,
\label{eq:simplemodel_homodyne}
\end{equation}
with $\sigma_{\rm H}=|g_0\rangle\langle e_+|+|g_0\rangle\langle e_-|$, and $|\psi(\tau)\rangle$ the state of the system at time $\tau$ after initiation by the trigger jump. The terms $\langle \psi(\tau)| \epsilon^*\sigma_{\rm H}|\psi(\tau)\rangle$ and $\langle \psi(\tau)| \sigma^\dagger_{\rm H}\epsilon|\psi(\tau)\rangle$ couple states $|e_{\pm}\rangle$ with $|g_0\rangle$, and due to the mapping of the ground-state coherence to the excited state, they oscillate as $\rho_{\rm g_{\pm},g_{0}}$. For $\epsilon$ sufficiently large these homodyne terms dominate. We then measure a second-order correlation function whose quantum beats oscillate at the Larmor frequency plus $\Delta_{\rm light}$.

\begin{figure}
\centering
\includegraphics[width=0.85\linewidth]{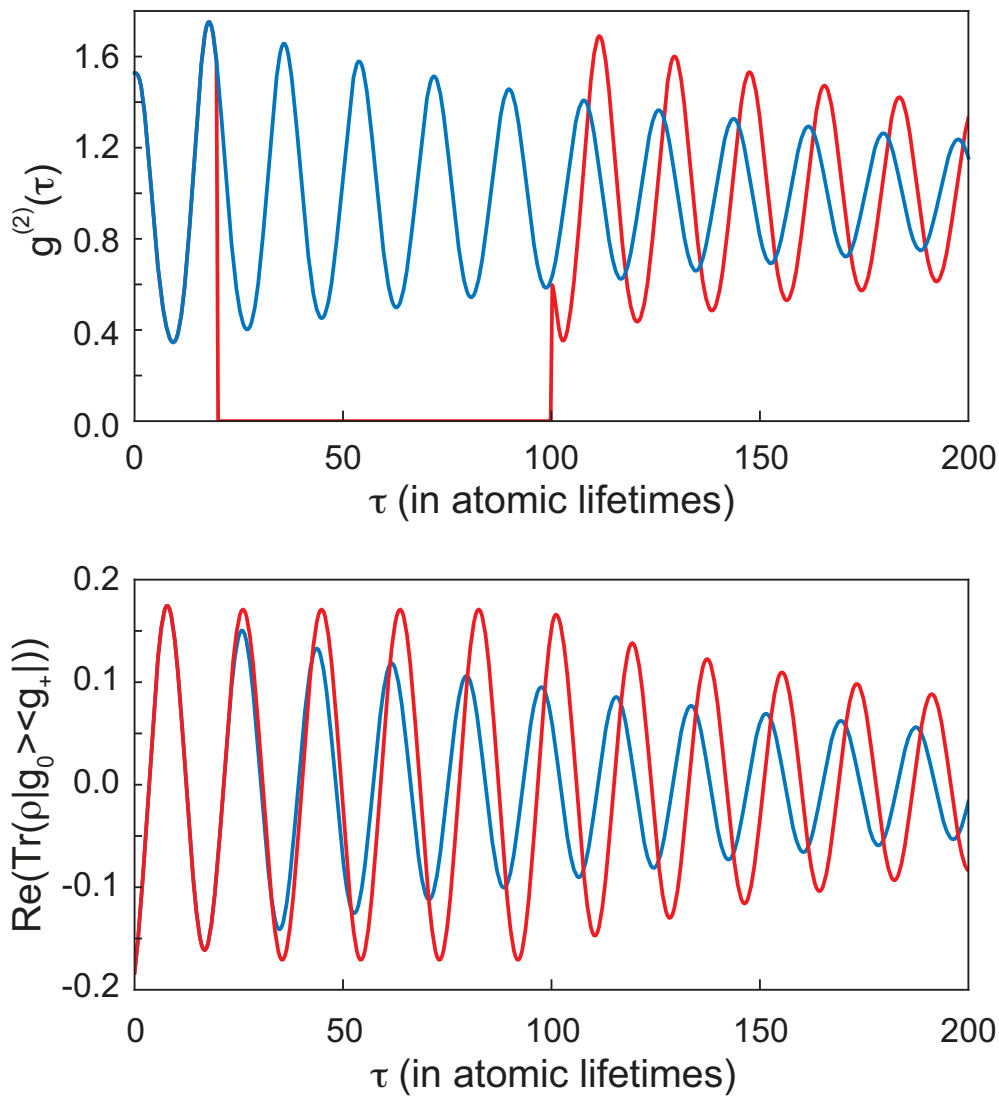}
\caption{Time response of the system when pulsing the drive (red) and with a continuous drive (blue). (a) Conditional intensity $g^{(2)}(\tau)$ (the traces initially overlap). (b) Atomic coherence between $|g_0\rangle$ and $|g_{+}\rangle$. For a drive strength corresponding to one photon in the driven mode, one fixed maximally coupled atom, and $\epsilon=0.1$.}
\label{fig:control_bb}
\end{figure}

The driving field can control the frequency shift and decoherence induced by Rayleigh scattering. Figure~\ref{fig:control_bb} shows an example of the proposed protocol, calculated on the basis of the simple model of a fixed atom and two cavity modes. Fig.~\ref{fig:control_bb}(a) displays the time evolution of the intensity correlation function of the undriven mode with no intervention (blue line) and with the drive laser ($\pi$ polarization) turned off 20 atomic lifetimes after the detection of the first photon (red line); the system then evolves freely until the drive is returned to
its previous value after a further 80 lifetimes (blue trace). Note that the amplitude of the red trace returns to the same value as before the turn off, while the phase accumulated is different in the presence of the drive and in the dark. We create and capture the coherence, preserving it in the dark, where it evolves without interrogation, and then we recover it. The phase difference visible after the oscillations return is a measure of
the average number of intervening quantum jumps. Figure~\ref{fig:control_bb}(b) shows how the ground-state coherence evolves in the dark for eighty atomic lifetimes, without any change in frequency or loss of amplitude due to Rayleigh scattering.

The model analyzed in this section is idealized as the considered transition has many more levels, and $\sigma$ and well as $\pi$ Rayleigh scattering occurs. It is nevertheless a good one to gain qualitative understanding of the phenomena. We have also developed a full quantum simulation of the problem including all relevant experimental realities. The details of that model and relevant calculations can be found in Ref.~\cite{norris12b}.

\section{Results}
\label{sec:results}
We now present the results of the feedback protocol: an increase in coherence time of the ground-state superposition, \emph{i.e.} an increase in beating amplitude, and the contrast in accumulated phase due to different precessing frequencies in the presence or absence of drive.

\begin{figure}[b]
\centering
\includegraphics[width=0.85\linewidth]{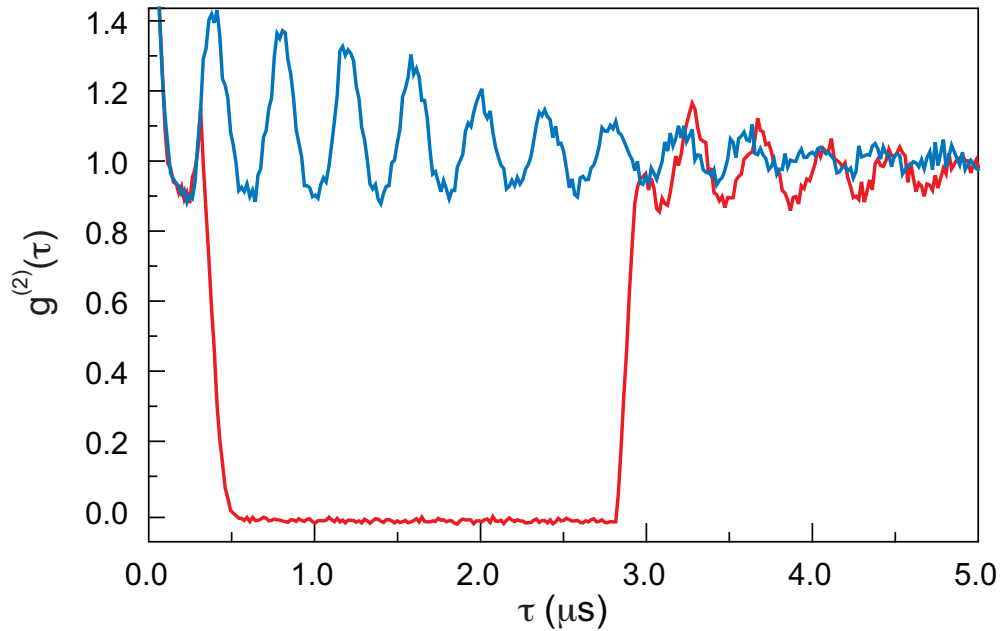}
\caption{Measured conditional intensity evolution, $g^{(2)}(\tau)$, of the undriven cavity mode in the presence of feedback (red) and with no feedback (blue). See text for a discussion of background suppression. For an effective atom number $N_{\rm eff}=0.4$, rotation of $1.2^\circ$ at HWP, and mean number of photons in the driven cavity mode of $n=1$.}
\label{fig:totaloff}
\end{figure}

Figure \ref{fig:totaloff} shows two experimental traces. Each pair of traces represents an average over a little more than 20 million photon arrivals. The time that elapses for each dataset, at an average rate of 60,000 counts/second (count rate of detectors A and B), is about $\sim$6 minutes. Each bin of 16 ns can have between 3000 and 4000 counts. The blue trace corresponds to no feedback pulse, and the red to a pulse of 2.5 $\mu$s, which is equivalent to 96 atomic lifetimes. The amplitude of the red trace is clearly larger when it returns, and there is a phase shift. Our two qualitative predictions are represented in the data.

The quantitative behavior depends on the exact mixing of the driving field and scattered light in the homodyne detection ($1.2\pm0.2^\circ$ at HWP), the number of photons in the driven mode ($n=1\pm0.3$), as well as details of the number ($N_{\rm eff}=0.4\pm0.2$), velocity ($10\leq v_p\leq15\mkern2mu{\rm ms}^{-1}$), and angular distribution ($\theta \leq 20$ mrad) of the atoms. At any given time many atoms are present in the cavity mode. This gives rise to a peak (bunching) in $g^{(2)}(0)$. The data shows a count rate with the ``off'' position greater than zero due to a background unrelated to light scattered from the cavity $V$ mode. The source of the background is scattering off the cavity mirrors of photons from the Magneto-Optical Trap laser beams and APD dark counts. These photons are uncorrelated and we set this level as the zero in the displayed experimental $g^{(2)}(\tau)$. The amount of background suppressed is the distance between the mark where $g^{(2)}(\tau)=0$ and the bottom line of the frame in the figure (about 0.05 in this figure).

We estimate the intra-cavity driven mode photon number using an independent calibration of the efficiency and the size of the signal when we mix the polarizations in the undriven mode. The result is a photon number of $n=1$ with an uncertainty of 30\%.

We repeat the measurements for different delay times so that we can extract the amplitude of the oscillation as a function of pulse width. Each trace taken has one side with no feedback and one with feedback. We perform a least squares fit between the no feedback case and the feedback case making adjustments to match the amplitude and phase of the oscillation after the drive returns to its steady state. This is done with an algorithm and allows us to determine an error for the amplitude scaling and the phase shift.

\begin{figure}
\centering
\includegraphics[width=0.85\linewidth]{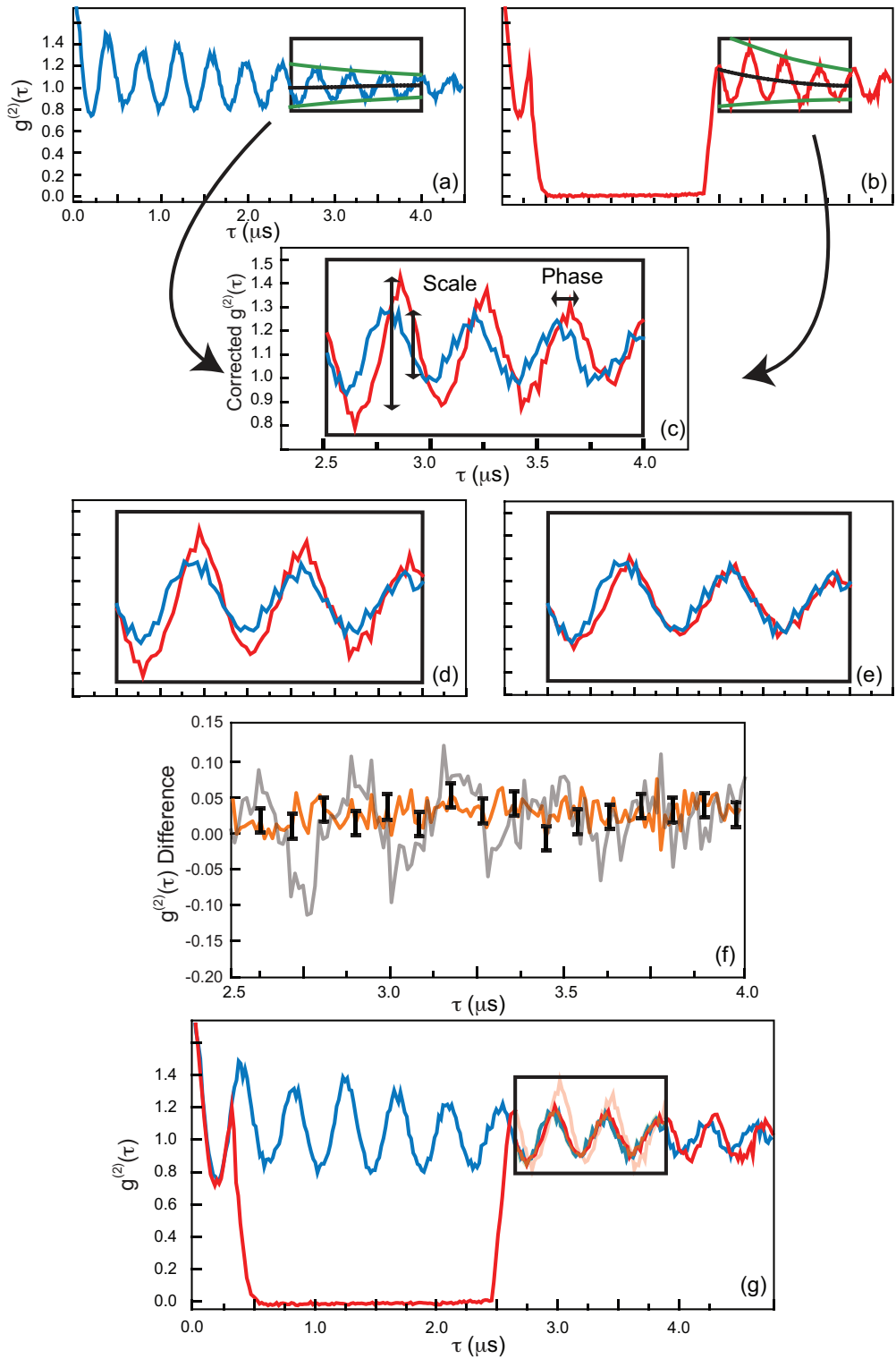}
\caption{Least-squares fitting process. (a)  $g^{(2)}(\tau)$  with no feedback and polynomial fits to maxima and minima in the restricted range. (b)  $g^{(2)}(\tau)$ with feedback and similar polynomial fits. (c) Fitting region after removing backgrounds to show phase and amplitude difference. (d) After shifting. (e) After scaling. (f) Difference between the traces before (gray) and after (orange) the fitting. (g) End result with the original trace for comparison.}
  \label{fig:fit_proc}
\end{figure}

Figure \ref{fig:fit_proc} explains the fitting process. First, we restrict the fitting to a limited range of the data [Fig.~\ref{fig:fit_proc}(a) and (b)]; for all data sets we use the time of the drive turn-on as the starting point and a fixed ending point at $4.7\mkern2mu\mu{\rm s}$. We then fit a second-order polynomial to the maxima and minima of the oscillations. We take an average of these curves and subtract them from the data [Fig.~\ref{fig:fit_proc}(c)]. This effectively removes differing backgrounds between the sides of the data. In the final step we perform the least-squares fit between the two batches, using two parameters: a time shift [Fig.~\ref{fig:fit_proc}(d)] and a scaling [Fig.~\ref{fig:fit_proc}(e)]; the residuals are shown in Fig.~\ref{fig:fit_proc}(f). We bin the data in $1.64\mkern2mu{\rm ns}$ ($16.4\mkern2mu{\rm ns}$) to optimize the phase shift (amplitude) extraction. Figure~\ref{fig:fit_proc}(g) illustrates the results of the process~\footnote{MATLAB script available at: \url{http://hdl.handle.net/1903/13307}}.

\begin{figure}
\centering
\includegraphics[width=0.95\linewidth]{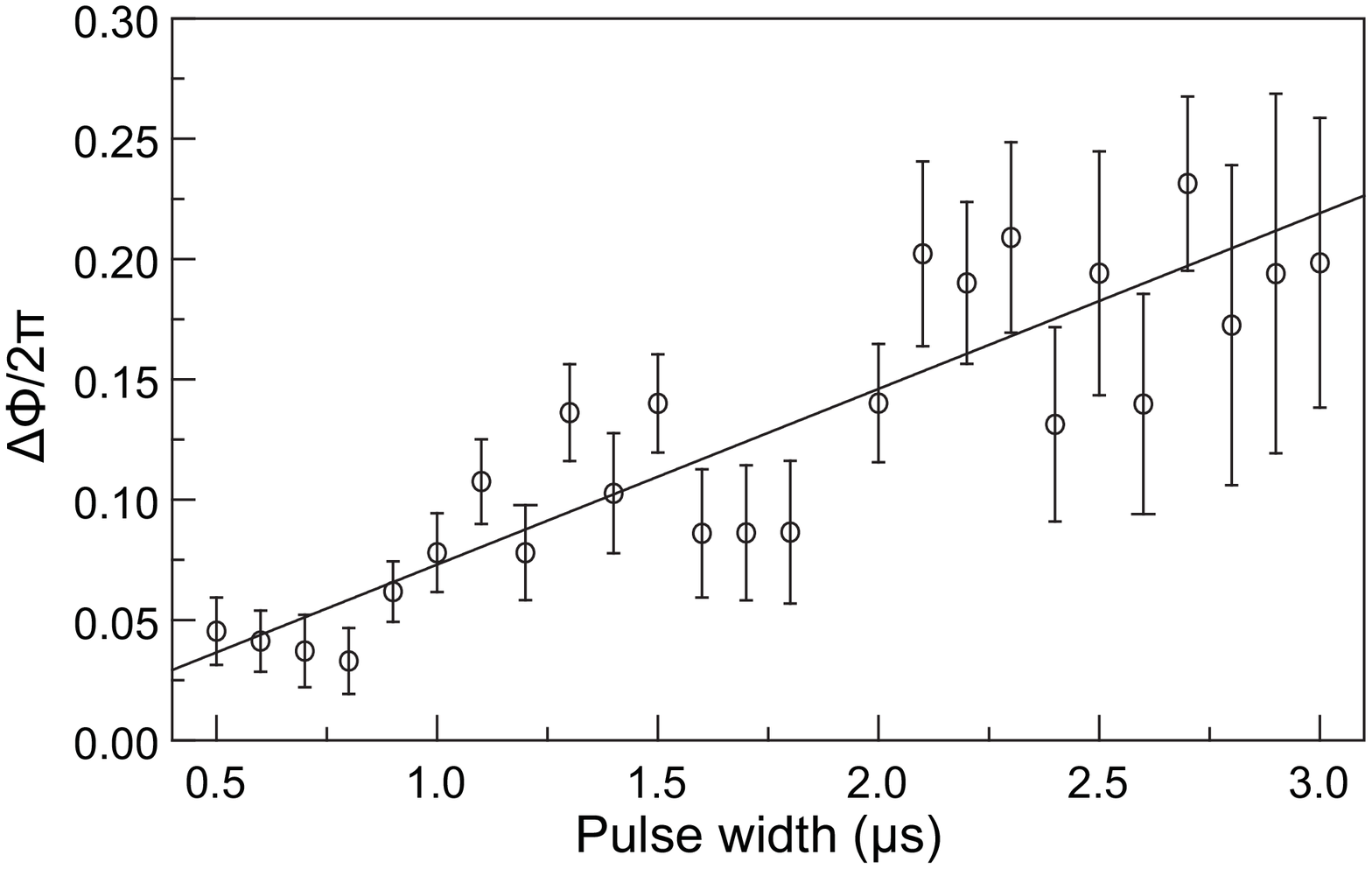}
\caption{Accumulated phase shift as a function of feedback pulse width. The continuous line is the least-squares fit to the data.}
\label{fig:phase}
\end{figure}

The fitting yields a phase shift and an amplitude change. First, we look at how the phase shift changes as the width of the feedback pulse is increased. The beat frequency is lowest (equal to the Larmor precession frequency) when the drive is off, and higher when it is on. The phase accumulated by the two $g^{(2)}(\tau)$ branches at a time $\tau$ after the feedback pulse is
\begin{align}
  \phi_- & = \omega_{\rm on}\tau,\\
  \phi_+ & = \omega_{\rm on}\tau_0 + \omega_{\rm off}(\tau_{\rm f} - \tau_0) + \omega_{\rm on}(\tau - \tau_{\rm f}),
\end{align}
for the phase accumulated without (-) and with (+) feedback, where $\tau_0$ is the time when the drive turns off and $\tau_{\rm f}$ is the time when the drive turns back on. We take the difference to obtain an expression for the phase shift:
\begin{equation}
\label{eq:phase-shift}
  \Delta \phi = \phi_- - \phi_+ = \left(\omega_{\rm on} - \omega_{\rm off}\right) (\tau_{\rm f} - \tau_0) = \Delta_{\rm light} (\tau_{\rm f} - \tau_0).
\end{equation}
The expression is linear in $\tau_{\rm f}$, and similar behavior is shown by the data in Fig.~\ref{fig:phase}. A least-squares fit (continuous line) yields a slope of $\Delta_{\rm light}/2\pi=0.073\pm0.004\mkern2mu{\rm MHz}$, consistent with the prediction from the simple model, Eq.~(\ref{eq:jumpshift}), of $\Delta_{\rm light}/2\pi=0.075\pm0.025\mkern2mu{\rm MHz}$ for $n=1\pm0.3$.

\begin{figure}
\centering
\includegraphics[width=0.95\linewidth]{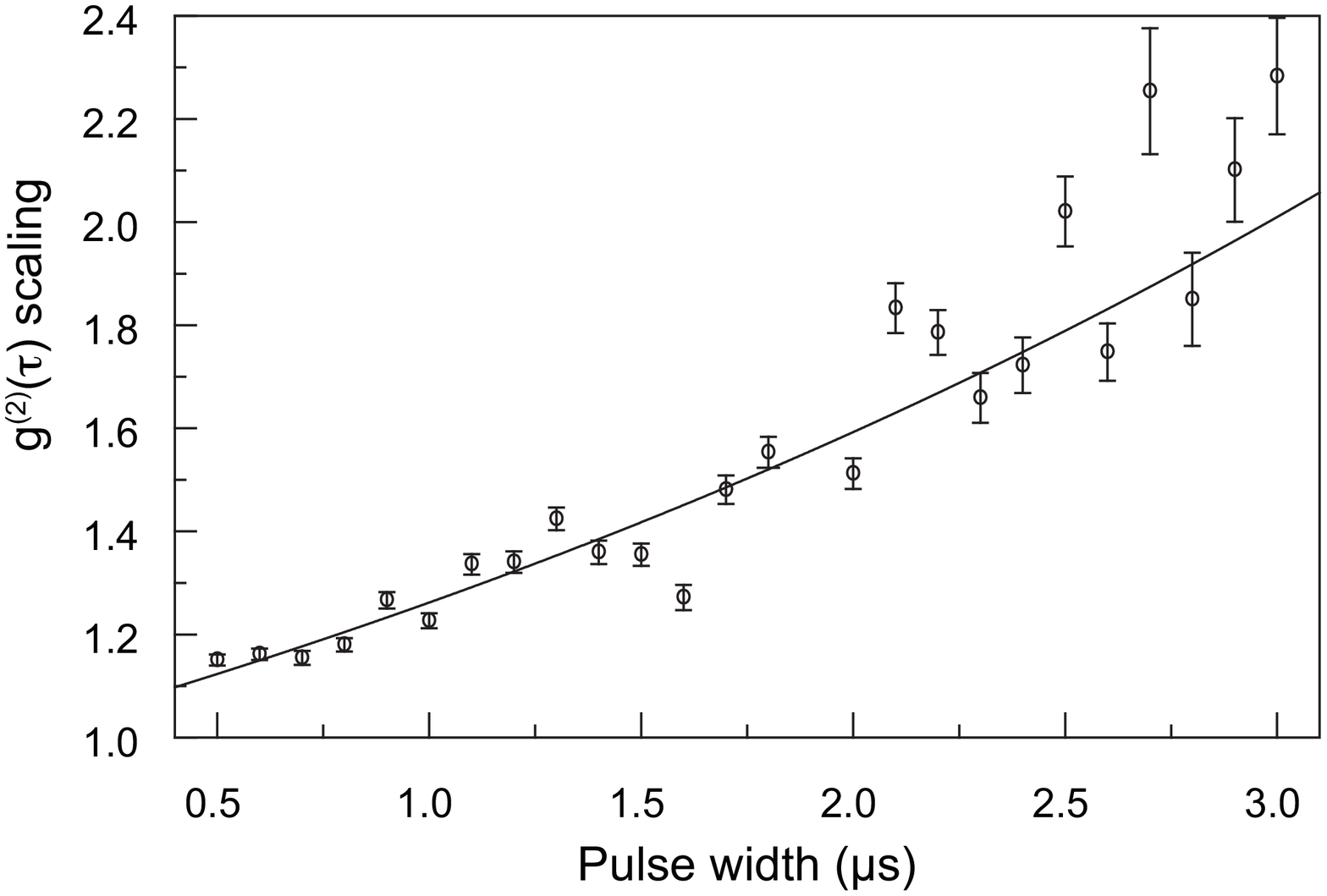}
\caption{Measured scale factor for the amplitude of the oscillations in $g^{(2)}(\tau)$ as a function of feedback pulse width. The continuous line is the expected result from Eq.~(\ref{eq:decohsimple}).}
\label{fig:yscale}
\end{figure}

Figure \ref{fig:yscale} presents results for the scaling of the oscillating in $g^{(2)}(\tau)$ as a function of feedback pulse width. This is a quantitative measure of the suppression of decoherence. If, for example, we let the system evolve in the dark for more than $2.5\mkern2mu\mu{\rm s}$, the amplitude of the oscillation is found to be a factor of two larger than without the feedback pulse; the protocol is clearly successful in suppressing decoherence. There other sources of decoherence, though, the dominant ones being the transit time of the atoms through the cavity mode and the angular distribution of their trajectories. We take the following simple additive model for decay of decoherence due to quantum jumps, rate $\Gamma_{\rm decoh}$, in the presence of other sources, rate $\Gamma_{\rm other}$:
\begin{equation}
 g^{(2)}(\tau)\propto1+e^{-\Gamma_{\rm other}\tau} e^{-\Gamma_{\rm decoh} \tau} \cos(\omega \tau).
 \label{eq:decohsimple}
 \end{equation}
It predicts the scaling $[g_{+}^{(2)}(\tau)-1]/[g_{-}^{(2)}(\tau)-1]=e^{\Gamma_{\rm decoh}(\tau_{\rm f} - \tau_0)}$, with $\tau_{\rm f} - \tau_0$ the feedback pulse width. The decoherence rate obtained from the data is $\Gamma_{\rm decoh}/2\pi=0.037\pm0.001\mkern2mu{\rm MHz}$, compared with an expected value of $\Gamma_{\rm decoh}/2\pi=0.032\pm0.010\mkern2mu{\rm MHz}$ from Eq.~(\ref{eq:jumpwidth}), at $n=1\pm0.3$.

\begin{figure}
\centering
\includegraphics[width=0.95\linewidth]{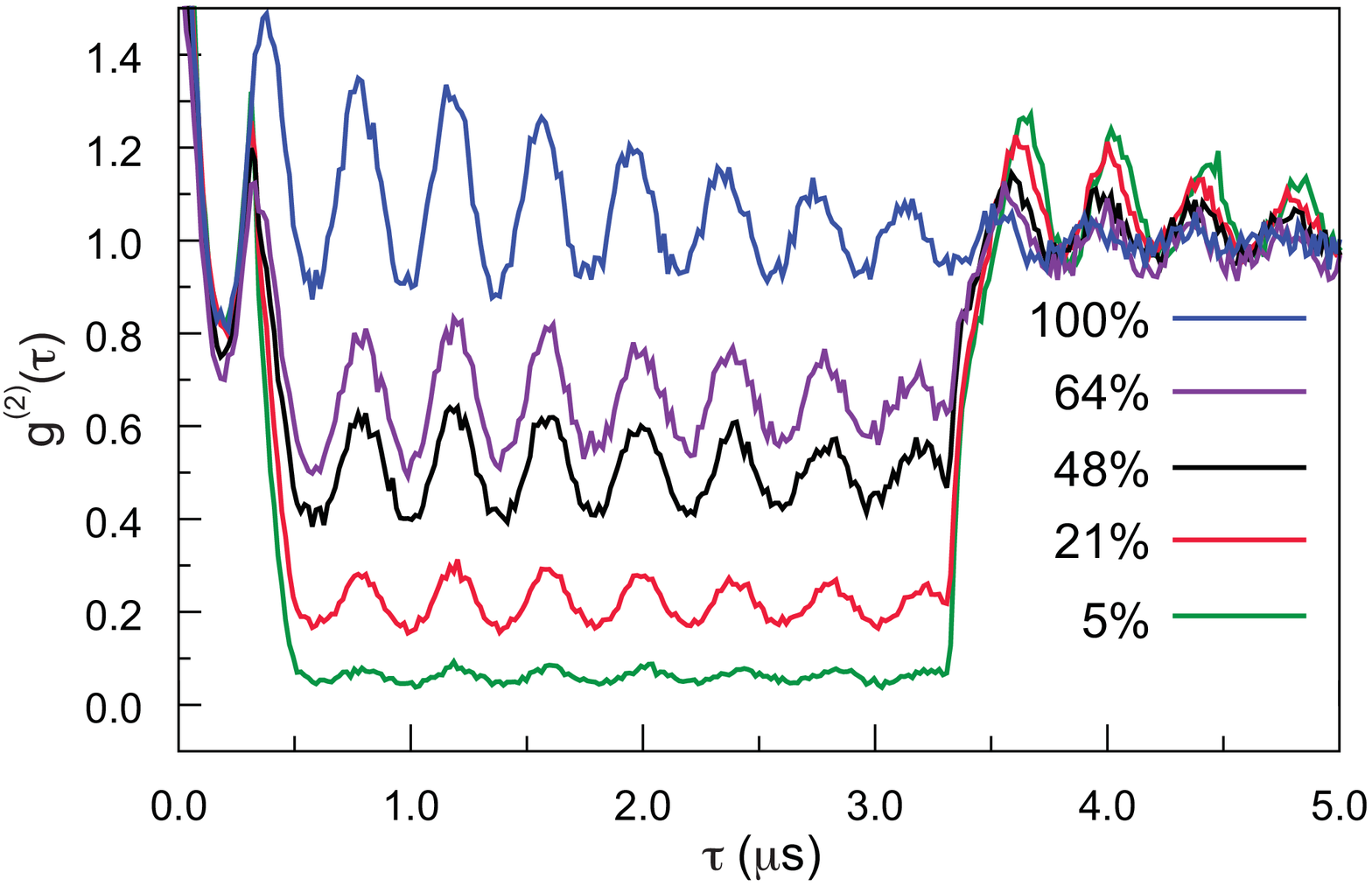}
\caption{Measured conditional intensity, $g^{(2)}(\tau)$, with fixed feedback pulse length ($3\mkern2mu\mu{\rm s}$) and variable amplitudes as indicated by the color code. For an effective atom number $N_{\rm eff}=0.4$, rotation of $1.2^\circ$ at HWP, and mean number of photons in the driven cavity mode of $n=1$.}
\label{fig:data_partial}
\end{figure}

A different way to carry out the investigation is to fix the feedback pulse width at $3\mkern2mu\mu{\rm s}$ and change the size of the drive, from the full drive (100\%) to smaller values, noting how the amplitude and phase of the oscillations change. Figure~\ref{fig:data_partial} presents a set of measurements with five different turn-off ratios. As the background suppressed in this figure is different for each trace, the distance between the mark for $g^{(2)}(\tau)=0$ and the bottom of the frame (about 0.1) shows the maximum amount we had to suppress. Once the drive returns to its starting level the changes in the amplitude and phase of the oscillations are significant, particularly for the 5\% case, and ordered according to color as we would expect.

\section{Discussion}
\label{sec:discussion}

\begin{figure}
\centering
\includegraphics[width=0.95\linewidth]{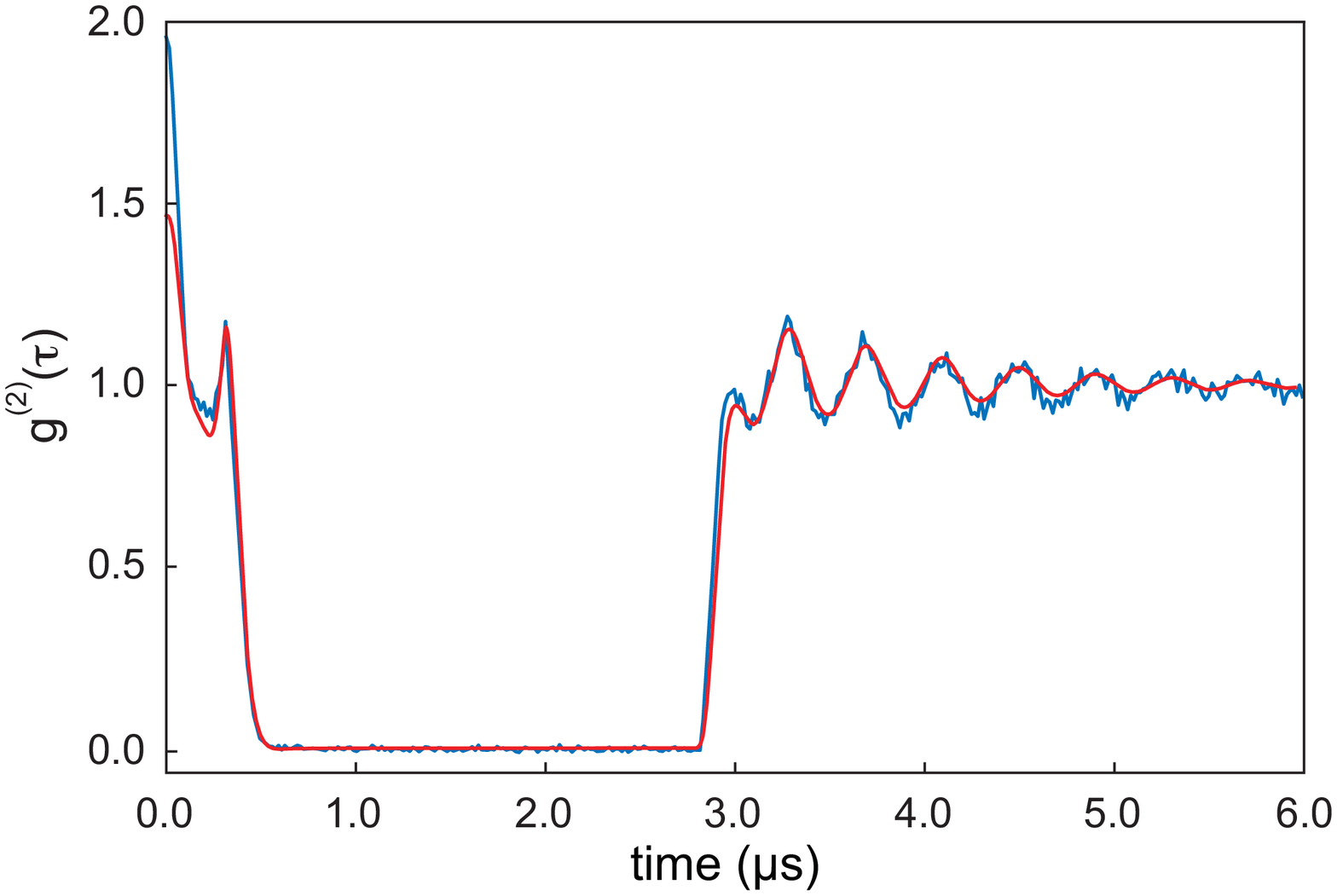}
\caption{Comparison of theory and experiment for controlled excitation. The fit uses effective atom number $N_{\rm eff}=0.55$, a rotation of $1.2^\circ$ at HWP, mean atomic speed $v_{\rm p}=13.5\mkern2mu{\rm ms}^{-1}$, a deviation of the atomic beam from perpendicular to the cavity axis $\theta=0.017\mkern2mu{\rm rad}$, and mean number of photons in the driven cavity mode $n=1.21$. }
\label{fig:comparisonbang}
\end{figure}

The comparison with theory starts with a numerical simulation of the experiment in the absence of feedback to obtain the best parameters for the effective number of atoms, number of photons in the driven mode, average atomic velocity, and the angle between the atomic beam and cavity axis. We make this fit after subtracting from the experimental data any background that prevents the signal from going to zero when the drive goes to zero, as is done for all the data figures in Sec.~\ref{sec:results}. We first adjust the background following a procedure similar to the first stage [Fig~\ref{fig:fit_proc}(a) and (b)] of the fitting process in Sec.~\ref{sec:results}. The amount of background suppression is equal to the distance between the mark for $g^{(2)}(\tau)=0$ and the bottom of the figure frame (about 0.05).

Using the fit parameters we calculate the controlled case. Fig.~\ref{fig:comparisonbang} shows an example of the results. The qualitative features of the data are all present in the model. Quantitatively, the model captures the phase shift and does an excellent job with the time constant of the cavity, which controls the decay and the rise of the signal when the pulse is applied. The difference between at $\tau=0$ may come from unaccounted contributions from multiple atoms and/or additional background. Modeling all decoherence processes is difficult; the model captures most of the decoherence, but makes a slight overestimate as the figure shows. The decoherence rate is very sensitive to the atomic velocity distribution, which is difficult to reproducibly control in the experiment to better than ten percent.

Considering comparisons with earlier work, the evolution of the ground state coherence takes place on time scales that allow implementation of feedback protocols with available laboratory equipment. This broadens the scope for experimental exploration compared to the hardware (time response) limitations of the quantum feedback previously implemented on the vacuum Rabi splitting \cite{smith02,reiner04a}. Working with this same ground state coherence we have shown in Ref.~\cite{norris11} how it is possible to modify the behavior by giving some time dependence to the drive. The combination of that idea with the current results points to new directions, which include the possibility of incorporating direct RF drives of the Larmor oscillation.

\section{Conclusions}
\label{sec:conclusion}

We have shown in this work the idea pioneered by Ramsey \cite{ramsey50} of letting quantum coherence evolve in the dark, is valid for conditional coherences, those not visible in the mean transmitted light and requiring the measurement of higher-order correlations for their study. Our use of feedback to counteract and measure the effects of Rayleigh scattering (both frequency shift and decoherence) shows that the qualitative behavior of our system is well understood, while we continue to better our quantitative understanding and detailed modeling. The reported protocol is simple, robust, and can improve the lifetime of a spontaneously generated coherence by a significant amount. It is an advance along the path to a new class of quantum feedback and control.

Work supported by CONACYT, M{\'e}xico, the NSF of the United States, and the Marsden Fund of the Royal Society of New Zealand.

\section{References}
\label{sec:refs}

\end{document}